\documentclass[pra,twocolumn,amsmath,amssymb,superscriptaddress]{revtex4-2}

\usepackage{graphicx,relsize,epstopdf,color,mathtools,bm,newtxtext,newtxmath,braket,rotating,dsfont,mathtools}

\usepackage[hyphenbreaks]{breakurl}
\usepackage[colorlinks=true,linkcolor=blue,citecolor=blue,urlcolor =blue]{hyperref}
\usepackage{float}
\usepackage{flafter}
\usepackage{placeins}
\usepackage{soul}
\usepackage{cancel}
\usepackage{comment}
\usepackage[table,xcdraw]{xcolor}
\newcommand{\Tr}{\mathop{\mathrm{Tr}} \nolimits}

\newcommand{\re}{\mathop{\mathrm{Re}} \nolimits}
\newcommand{\im}{\mathop{\mathrm{Im}} \nolimits}
\newcommand{\iu}{\text{i}}
\newcommand{\eu}{\text{e}}

\newcommand{\matriz}[1]{\mathsf{#1}}

\begin{document}

\title{Quantum multiphase estimation}

\author{M. Barbieri}
    \affiliation{Dipartimento di Scienze, Università degli Studi Roma Tre, Rome, Italy}
    \affiliation{ Istituto Nazionale di Ottica - CNR, Florence, Italy} 

\author{I. Gianani}
    \affiliation{Dipartimento di Scienze, Università degli Studi Roma Tre, Rome, Italy}

\author{A.~Z.~Goldberg}
    \affiliation{National Research Council of Canada, 100 Sussex Drive, Ottawa, Ontario K1N 5A2, Canada}

\author{L.~L. S\'anchez-Soto}
    \affiliation{Departamento de \'Optica, Facultad de F\'{\i}sica, Universidad Complutense, 28040~Madrid, Spain}
    \affiliation{Institute for Quantum Studies, Chapman University, Orange, CA 92866, USA}
    \affiliation{Max-Planck-Institut f\"ur die Physik des Lichts, 91058 Erlangen, Germany}

\begin{abstract}
Quantum phase estimation is fundamental to advancing quantum science and technology. While much of the research has concentrated on estimating a single phase, the simultaneous estimation of multiple phases can yield significantly enhanced sensitivities when using specially tailored input quantum states. This work reviews recent theoretical and experimental advancements in the parallel estimation of multiple arbitrary phases. We highlight strategies for constructing optimal measurement protocols and discuss the experimental platforms best suited for implementing these techniques. 
\end{abstract}

\maketitle

\section{Introduction}

Quantum metrology has attracted significant interest in recent years for its capability to estimate unknown parameters with a precision beyond the reach of classical methods~\cite{Paris:2009aa,Giovannetti:2011aa,Demkowicz:2015ab,Sidhu:2020aa,Polino:2020aa,Barbieri:2022aa,Alodjants:2024aa}. Research, both theoretical and experimental, has largely centered on the estimation of a single phase--a key paradigm, especially in photonic systems, which are the leading platform for these studies. For single-parameter estimation, the ultimate sensitivity bounds and explicit conditions for their saturation are well known~\cite{Braunstein:1994aa,Braunstein:1996aa}. In this scenario, classical methods are bound by the standard quantum limit (SQL)~\cite{Braginski:1975aa}, which dictates that the estimation error scales as $\nu^{-1/2}$, where $\nu$ is the number of independent measurements performed. However, by leveraging quantum resources, this limit can be overcome, achieving a higher precision that follows the Heisenberg limit (HL), where the error scales more favorably as $\nu^{-1}$~\cite{Giovannetti:2004aa,Giovannetti:2006aa,Demkowicz:2012aa}.

Several proof-of-principle experiments have demonstrated phase estimation below the SQL limit in various systems, such as interferometers~\cite{Xiao:1987aa,Grangier:1987aa,Slussarenko:2017aa,Pradyumna:2020aa,Cimini:2023aa,Kalinin:2023aa}, atomic clocks~\cite{Leroux:2010aa,Hosten:2016aa}, and magnetometers~\cite{Wolfgramm:2010aa,Wasilewski:2010aa,Horrom:2012aa,Sewell:2012aa}. This quantum enhancement has been strikingly demonstrated in the detection of gravitational waves~\cite{Aasi:2013aa,Tse:2019aa} and is in the realm of modern quantum sensing~\cite{Degen:2017ab,Aslam:2023aa}.

However, a substantial class of problems resists efficient framing as the estimation of a single parameter. This multiparameter scenario, which involves the simultaneous estimation of two or more parameters~\cite{Szczykulska:2016aa,Albarelli:2020aa,Demkowicz:2020aa,Belliardo:2021aa}, represents a relatively recent frontier that holds significant potential for diverse sensing applications, such as measuring inherently vector quantities, including fields~\cite{Koschorreck:2011aa,Baumgratz:2016aa,Altenburg:2017aa,Apellaniz:2018aa,Kaubruegger:2023aa}, displacements~\cite{Genoni:2013aa,Hanamura:2021aa,Agarwal:2022aa}, and rotations~\cite{Goldberg:2018aa,Goldberg:2021aa,Eriksson:2023aa}. Recent advances have concluded that estimating multiple parameters simultaneously may achieve greater sensitivity than estimating them individually~\cite{Gorecki:2022aa}, and have focused on the roles of entanglement~\cite{Ciampini:2016aa}, as well as the impact of noise~\cite{Escher:2011aa,Yue:2014aa,Albarelli:2018aa,Zhou:2023aa}. Specific examples have been studied, encompassing measurement strategies for problems including measurement on biological systems~\cite{Taylor:2016aa}, phase imaging~\cite{Tsang:2016aa,Rehacek:2017ab,Rehacek:2017aa}, spectroscopy and frequency measurements~\cite{Ansari:2021aa,Mazelanik:2022aa,Krokosz:2024aa}, quantum sensing networks~\cite{Proctor:2018aa,Ge:2018aa,Gessner:2020aa,Kim:2024aa}, joint estimation of phase and loss~\cite{Pinel:2013aa,Crowley:2014aa}, phase and phase diffusion~\cite{Vidrighin:2014aa,Altorio:2015aa,Szczykulska:2017aa,Jayakumar:2024aa}, phase and indistinguishability~\cite{Knoll:2023aa}, and estimation tasks with incomplete knowledge of the measurement device itself~\cite{Altorio:2016aa}.

However, in this multiparameter scenario there are still several open questions. For instance, while the single-parameter case  is well understood, few recipes to saturate the ultimate bounds are known in the multiparameter case~\cite{Helstrom:1974ab,Matsumoto:2002ab,Ragy:2016ab,Yang:2019aa,Sidhu:2021aa,Chen:2024aa}. Here, due to possible noncommutativity of the quantum measurements required to simultaneously optimize the estimation of different parameters, it may not be possible to optimally estimate all parameters at the same time.  The proper definition of general quantum bounds still requires additional investigations.

To effectively investigate these multiparameter tasks, it is essential to identify a specific scenario and a corresponding experimental platform. The multiphase problem, where the parameters to be estimated consist of a set of optical phases, offers such a scenario. Numerous theoretical studies have explored this direction~\cite{Macchiavello:2003aa,Spagnolo:2012aa,Humphreys:2013aa,Liu:2016aa,Gagatsos:2016aa,Zhang:2017aa,Goldberg:2020ad,Hong:2021aa,Gebhart:2021aa}. Recent findings have established necessary and sufficient conditions for determining the optimal projective measurements for pure states~\cite{Pezze:2017aa}. Additionally, advancements have been made in formulating generalized matrix bounds~\cite{Conlon:2024aa} and identifying optimal states~\cite{Gessner:2018aa}.  In this paper, we review these capabilities as well as the latest experimental implementations.

\section{Background} 

Light is often used as the probe in sensing tasks. Perhaps the simplest case is the detection of specimens by absorption~\cite{McDonagh:2008qy}: a specimen's distinct features allow one to identify the presence of a compound by looking at the absorption spectrum. Depending on the target compound, it may be more convenient to detect its fluorescence when pumped with light at shorter wavelengths. 

A different approach employs an interferometric technique. Here, the medium is not directly probed, as for the spectroscopic approach; instead, its properties are inferred by observing the relative phase in a Mach-Zehnder interferometer, or in an equivalent arrangement. One of the arms serves as a reference, while the second either contains or is in close contact with the target, so that the optical path can be modified by parameters such as concentration of a chemical species, temperature, pressure, and so forth. This approach finds its justification in the exceptional sensitivity that interferometers can deliver, being patently affected by changes of the optical path as small as a fraction of a wavelength. Sensing thus occurs based on a transduction mechanism, in which the quantity of interest influences the optical phase and, in turn, this is the quantity actually being accessed. 

\begin{figure}
    \centering
    \includegraphics[width=0.85\columnwidth]{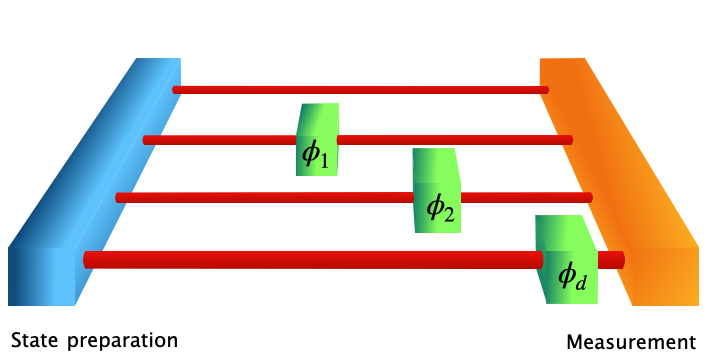}
    \caption{Conceptual scheme of all multiphase estimation protocols.}
    \label{fig:concept}
\end{figure}

This concept can be generalised to sensors comprised of multiple arms, each with its own phase. The general scheme is illustrated in Fig.~\ref{fig:concept}: a set of $d$ phases, $\phi_1$, $\phi_2$,$\dots$,$\phi_d$, are placed in a $(d+1)$-arm interferometer, the first arm serving as a reference for the others. The probing light is prepared in an initial state, determined by the properties of the source and on the specific arrangement that distributes it across the $d+1$ modes. Finally, these modes are recombined and sent to detectors. This scheme forms the template for a single, compact device able to sense a quantity in different locations at once, thus enabling, for instance, the evaluation of gradients, and, in general, adding to the capability of single-phase sensors.
This is quite different from an array of multiple single-phase sensors, in that the probe preserves its coherence when distributed, while a sensor array can only compare the individual outputs.

The same concept also serves as a model for phase-contrast imaging. By this technique it is possible to obtain an image of a nearly transparent object based on the optical phase it imparts. Although the technical requirements are vastly different from those of a pressure gradient sensor, to quote but one example, the basic features of a phase microscope are still captured by the scheme in Fig.~\ref{fig:concept}.

As a final example, phase objects can be dislocated across a network, thus realising a distributed sensing task, relevant for seismic monitoring~\cite{Schenato:2017uq}. This represents a more involved instance, as one should account for realistic limitations in measuring capabilities and the need for communication across nodes; however, the multiphase setting is able to provide guidance and insight on many fundamental aspects.

In the classical picture, the phase $\phi$ captures the amount of oscillation of an electromagnetic field and serves as a key parameter for probing the properties of the medium through which the field propagates. Its status as a genuine quantum observable has been vigorously debated~\cite{Pegg:1989aa,Lynch:1995aa,Perinova:1998aa,Luis:2000aa}. To elucidate this issue, we consider a single-mode field,  equivalent to a quantum harmonic oscillator described by annihilation ($a$) and creation ($a^\dagger$) operators, with the basic commutation relation $[a, a^\dagger] = 1$. The phase $\phi$ naturally emerges as the conjugate variable to the number operator $N = a^\dagger a$, as encapsulated by the Dirac commutation relation $[\phi, N]= i$~\cite{Dirac:1927aa}. This, in turn, suggests the number-phase uncertainty relation
\begin{equation}
\mathrm{Var} (N)\,  \mathrm{Var}(\phi) \ge \frac{1}{4} \, , 
\end{equation} 
in terms of the corresponding variances. This is now known to be flawed~\cite{Susskind:1964aa,Carruthers:1968aa}, prompting various alternative formulations.

For our present discussion, this subtlety remains marginal, as the phase is not directly measured, but rather inferred as a parameter governing observed outcomes. Operationally, it appears in the unitary transformation $U(\phi) = \exp(\iu \phi N)$ generated by the  the photon-number operator $N$. To illustrate this, we consider the quantum state  most akin to a classical oscillating field: the coherent state $|\alpha\rangle$~\cite{Glauber:1963aa}. Defined by the eigenvalue equation $a | \alpha \rangle = \alpha | \alpha \rangle $, it is characterized by a complex amplitude  $\alpha$  encoding both field intensity and phase. Coherent states exhibit minimum uncertainty,  remain stable under free evolution,  and can be generated from the vacuum using the displacement operator  $D(\alpha) = \exp(\alpha a^{\dagger} - \alpha^{\ast} a)$.  Under the transformation $U(\phi)$, the coherent state evolves as $|\alpha\rangle \mapsto |\alpha \eu^{\iu\phi}\rangle$, recovering the expected classical evolution of field amplitudes $\langle {a}\rangle\mapsto \eu^{\iu \phi}\langle {a} \rangle$. However, this transformation does not provide a direct means to measure the phase.  

How can it be measured, then?  Not with a detector that merely records the probabilities of finding a certain number of photons, as the phase information in $|\alpha \eu^{\iu \phi}\rangle$ is exclusively in the phase differences between the subspaces with different numbers of photons. Rather, light must be compared to some reference state, with the phase information arising in an interference pattern resulting from this comparison. A standard technique to do so is homodyne detection, where the light is interfered with another coherent state $|\beta\rangle$ at a balanced beam splitter and then the intensity difference between the two output ports is registered. This amounts to a measurement of the operator $\iu\beta  a^{\dagger} -\iu  \beta^\ast a$ on the system of interest, which, for a coherent state $|\alpha\rangle$, yields $2 \im(\alpha\beta^\ast)$. Applying a known phase to the reference state $|\beta\rangle \mapsto |\beta\eu^{\iu\theta}\rangle$ allows a sinusoidal inteference pattern to build up with $\theta$: $2\im(\alpha\beta^\ast \eu^{-\iu \theta})=2|\alpha\beta|\sin(\arg\alpha-\arg\beta-\theta)$. When $\beta$'s phase is known, this is a direct method for finding the phase of $\alpha$; otherwise, only the \textit{relative} phase $\arg\alpha-\arg\beta$ can be discerned. 

The treatment extends in a straightforward manner to multiple phases associated to orthogonal modes; i.e, modes associated to destruction operators $a_i$ and $a_j$ satisfying $[a_i,a^\dagger_j]=\delta_{ij}$. In this case, the evolution occurs independently on each mode $|\psi\rangle\to \eu^{\iu\sum_j \phi_j N_j} |\psi\rangle$. There remains the fact that each phase can be retrieved only relative to a reference mode.

To obtain the sought high-level description of the sensor, we adopt the framework of local estimation theory: we identify a vector of parameters of interest $\{\phi_1,\dots,\phi_d\}$, whose value is assessed by a series of $\nu$ runs of sensing. We will not gain access to their true values, but rather to their estimates, which we denote as  $\{\hat{\phi}_1,\dots,\hat{\phi}_d\}$, whose uncertainty should be minimised within the limits imposed by the number of repetitions and by the intensity of the light probe. In order to obtain a quantitative figure of merit, we introduce the \emph{covariance matrix} $\matriz{V}$, whose elements are defined as  
\begin{equation}
    \label{eq:covfefe}
    \matriz{V}_{ij} = \langle \hat{\phi}_i \hat{\phi}_j\rangle-\langle \hat{\phi}_i \rangle \langle \hat{\phi}_j\rangle,
\end{equation} 
where the angular brackets denote an average over the distribution of the parameters of interest. In particular, its diagonal elements are the variances on the individual phases. The estimated values may deviate from the true values due to finite statistics, but also due to the presence of a bias in the estimation; here we consider \emph{unbiased estimators} for which this latter contribution vanishes.

\section{Quantum phase estimation}

In this section, we briefly explore how quantum theory places bounds on the ultimate precision with which a single phase can be measured.  

It is imperative to compare various probe states and measurements thereon to find the optimal method for probing and estimating the phase. Given a probe state upon which one unknown parameter is  imprinted, quantum estimation theory provides a quantity known as the quantum Fisher information (QFI) that encapsulates the ultimate precision with which the parameters may be estimated, optimized over all measurement strategies. 

The goals of quantum estimation theory are thus to a) identify probe states with the most QFI, b) devise measurement strategies that extract the most possible information about parameter-dependent states, and c) find paradigms where these results outperform the best possible results that can be achieved without consideration of quantum estimation. 

In this context, a phase $\phi$ is imparted on any probe state $|\psi_0\rangle$ via a unitary transformation $|\psi_0 \rangle \mapsto|\psi_\phi\rangle=U(\phi)|\psi_0\rangle$, as discussed before. Quantum theory dictates that any measurement on a quantum state is, in general, described by a positive operator-valued measure (POVM)~\cite{Holevo:2003fv}, a more general framework than the traditional von Neumann projective measurements.  A POVM consists of a set $\{ \Pi_{n} \}$ of self-adjoint  ($\Pi_n^\dagger=\Pi_n$) and positive semidefinite ($\Pi_n \succeq 0$) operators that  satisfy the completeness condition $\sum_{n} \Pi_{n} = \openone$.  Each $\Pi_{n}$ corresponds to a possible measurement outcome, labeled here by a discrete index (which could also be continuous, replacing sums with integrals). The probability of obtaining outcome $n$  when measuring a quantum state $\varrho$ follows the Born rule~\cite{Peres:2002oz}
\begin{equation}
p(n | \phi)= \Tr (\varrho_\phi \Pi_n) \, .
\label{eq:Born}
\end{equation} 
Unlike projective measurements, POVMs do not require  measurement operators to be orthogonal projectors ($\Pi_{n}^{2} \neq \Pi_{n}$, in general). They allow for more measurement outcomes than the Hilbert space dimension, extracting richer information while minimizing disturbance to the system. This makes POVMs particularly useful in realistic, noisy environments where standard measurements may be inadequate or suboptimal.

From the probabilities \eqref{eq:Born}, the Fisher information (FI) can be calculated as~\cite{Fisher:1925aa}
\begin{equation}
    F_{\Pi}(\phi )=\sum_n p(n |\phi )\left [\frac{\partial \ln p(n | \phi)}{\partial \phi}\right ]^2
    \label{eq:FI classical}
\end{equation}
and its maximization over all measurement strategies gives the QFI:
\begin{equation}
    Q_\varrho(\phi)=\sup_{\{\Pi_n\}} F_{\Pi}(\phi)= 4 \,  (\langle \partial_\phi \psi_\phi|\partial_\phi \psi_\phi\rangle-|\langle  \psi_\phi|\partial_\phi \psi_\phi\rangle|^2 ),
    \label{eq:QFI single param deriv}
\end{equation} 
where the final equality holds only for pure states and we use the notation $|\partial_\phi \psi_\phi\rangle= \partial | \psi_\phi\rangle/\partial \phi$. The QFI $Q_{\varrho}(\phi)$ is a property of the quantum state $\varrho_{\phi}$ only, since it derives from an optimization over all possible measurements.

Intuitively, the FI  quantifies the sensitivity of a system to changes in $\phi$: a larger amount of information corresponds to greater variations in the output probabilities. Geometrically, FI defines a curvature on the statistical manifold of probability distributions, determining how distinguishable different parameterized distributions are. This implies that the more curved (i.e., informative) the parameter space is, the more precisely the parameter can be estimated. The FI is also directly related to the statistical distance between two probability distributions. Specifically, for two neighboring distributions, $p(n|\phi)$ and $p(n|\phi + d\phi)$, the infinitesimal statistical distance is given by
$ds^{2} = F(\phi) d\phi^{2}$. This relation highlights how FI governs the local geometry of the parameter space, setting a natural scale for parameter estimation.

All of the above is useful because of the Cram\'er-Rao bound~\cite{Cramer:1946aa,Rao:1945aa} and its quantum counterpart, which dictate that
\begin{equation}
    \mathrm{Var}(\hat{\phi})\geq \frac{1}{F_{\Pi}(\phi )} \geq 
    \frac{1}{Q_\varrho(\phi)}.
    \label{eq:CRB single param}
\end{equation} 
This means that the lowest uncertainty on an estimate of $\phi$ will be obtained by using a probe state with the largest QFI $Q_\varrho(\phi)$ and an optimal measurement for that state. The QFI has numerous properties, including convexity properties that ensure pure probe states to be optimal, and is conveniently rewritten in terms of the generator $N$ of the transformation $U(\phi)$ as :
\begin{equation}
    Q_\varrho(\phi)=4 \, \mathrm{Var}_{\psi_{0}} (N) .
    \label{eq:QFI single param generator}
\end{equation} 
For single-parameter estimation of unitary transformations, the QFI is independent from the value of the parameter $\phi$ and there is always a measurement whose FI attains the QFI to saturate the rightmost bound in Eq.~\eqref{eq:CRB single param}. The leftmost bound in Eq.~\eqref{eq:CRB single param}, in turn, is guaranteed by classical statistics to be saturable in the asymptotic limit of many repeated trials, where the FI and QFI each get multiplied by the number of trials $\nu$ and the variance shrinks accordingly.

We can now establish the classical benchmark: sensing with coherent states. We identify $|\psi_0\rangle=|\alpha\rangle$ and $|\psi_\phi\rangle=|\alpha \eu^{\iu \phi}\rangle$, getting for the variance $Q_\alpha(\phi)=4 \, \mathrm{Var}_\alpha(N)= 4|\alpha|^2$. The minimum uncertainty thus scales as the inverse of the energy of the probe state $\bar{n} =|\alpha|^2$. Here, the optimal measurement strategy is  homodyne measurement by interfering the state $|\alpha \eu^{\iu\phi}\rangle$ with a reference state of known phase $|\beta\eu^{\iu \theta}\rangle$ and looking at the intensity distribution that oscillates as $2|\alpha\beta|\sin(\arg \alpha+\phi-\arg\beta-\theta)$, where $\arg \alpha-\arg\beta-\theta$ is known and $\theta$ is experimentally tunable. A propagation of errors from this intensity to a measurement of $\phi$ yields
\begin{equation}
    \mathrm{Var}(\hat{\phi})\geq \frac{\frac{1}{|\alpha|^2}+\frac{1}{|\beta|^2}}{4\cos^2(\arg \alpha+\phi-\arg\beta-\theta)}\geq {\frac{1}{4|\alpha|^2}+\frac{1}{4|\beta|^2}}.
\end{equation} 
When the reference beam's energy gets large ($|\beta|^2\to \infty$), this approaches the QFI for a judicious choice of $\theta$. This is what the QFI presupposes: complete and perfect ideal measurements with infinite external resources may be necessary to achieve the QFI's ``optimal'' measurement. Practical considerations are thus always necessary when finding optimal measurements, especially in quantum estimation, and this particular dependence on the presence of a phase reference has required correction multiple times (consider the case where the reference beam is split from the probe state to be sent down another arm of an interferometer; then $|\alpha|=|\beta|$ and one recovers half of the FI and twice the variance that one could have expected from the QFI of $|\alpha\rangle$ alone~\cite{MarcinDemkowiczDobrzanski2012}).

According to Eq.~\ref{eq:QFI single param generator}, the optimal probe is thus a state that maximizes the variance of $N$. The optimal probe state is given by a superposition of eigenstates of $N$ that respectively have maximal and minimal eigenvalues~\cite{Giovannetti:2011aa}: if $N$ is the upper limit on the number of photons in a state, this yield the optimal state $|\psi_{\mathrm{opt}} \rangle = (|0\rangle+|N\rangle)/\sqrt{2}$ for Fock states with $0$ and $N$ photons, respectively. The QFI for this state is $N^2$, which is four times the square of the state's average energy $\bar{n}=N/2$. For this state, we thus have
\begin{equation}
    \mathrm{Var}(\hat{\phi})\geq\frac{1}{4\bar{n}^2}.
\end{equation} 
This rapidly decreasing variance with energy of the probe is known as the HL and drops much faster than the coherent states' scaling known as the SQL.

The existence of a state of the form of $|0\rangle+|N\rangle$ presupposes the existence of a strong reference state; otherwise, the superposition should be treated as a probabilistic mixture of states $|0\rangle\langle 0|$ and $|N\rangle\langle N|$. We next turn to quantum-advantage states that have their phase references inbuilt. The famous ``NOON'' states are two-mode states of the form~\cite{Dowling:2008aa}
\begin{equation}
    |\psi_{\mathrm{NOON}}\rangle=\tfrac{1}{\sqrt{2}} 
    (|N, 0\rangle+|0, N\rangle ) .
\end{equation} 
One can consider an interferometer of other setup, where the first mode is subject to a unitary evolution $\eu^{\iu \phi N}$ while the second is not. The QFI for this case is again $N^2$ and now we do not need an external phase reference for measurement but, rather, projections onto states like $|\psi_{\mathrm{NOON}}\rangle$.

\section{Quantum multiphase estimation}

In the multiphase scenario, $d$ modes acquire $d$ phases via the unitary $U(\boldsymbol{\phi})=\exp (\iu \sum_{i=1}^d \phi_i N_i )$. How well can all of these phases be measured?

Many of the single-parameter estimation formulas generalize simply to the multiparameter case, with a few crucial caveats. First, the FI generalizes to a matrix with components similar to those in Eq.~\eqref{eq:FI classical}:
\begin{equation}
\label{eq:fim}
    \mathsf{F}_{ij}(\boldsymbol{\phi})=\sum_n p(n|\boldsymbol{\phi})\frac{\partial \ln p(n|\boldsymbol{\phi})}{\partial \phi_i}\frac{\partial \ln p(n|\boldsymbol{\phi})}{\partial \phi_j},
\end{equation} 
where we have dropped the subscript corresponding to the POVM. Second, the QFI generalizes to a matrix with components similar to those of Eqs.~\eqref{eq:QFI single param deriv}~and~\eqref{eq:QFI single param generator}~\cite{Paris:2009aa}:
\begin{align}
    \mathsf{Q}_{ij}(\boldsymbol{\phi})& =4 \re (\langle \partial_{\phi_i} \psi_{\boldsymbol{\phi}}|\partial_{\phi_j} \psi_{\boldsymbol{\phi}}\rangle-\langle  \psi_{\boldsymbol{\phi}}|\partial_{\phi_i} \psi_{\boldsymbol{\phi}}\rangle\langle  \partial_{\phi_j}\psi_{\boldsymbol{\phi}}| \psi_{\boldsymbol{\phi}}\rangle ) \nonumber \\
    & =4\, \mathrm{Cov}_\psi(N_i,N_j)   \label{eq:QFI multiparam pure}
\end{align}
for the symmetrized covariances
$\mathrm{Cov}_\psi(N_i,N_j)= \tfrac{1}{2}\langle \psi_0|{N_i N_j+N_jN_i}|\psi_0\rangle-\langle \psi_0|N_i|\psi_0\rangle\langle \psi_0|N_j|\psi_0\rangle$. Together, these comprise a multiparameter (quantum) Cram\'er-Rao bound (QCRB) for the covariance matrix from Eq.~\eqref{eq:covfefe}:
\begin{equation}
    \matriz{V} \succeq \matriz{F}^{-1}\succeq \matriz{Q}^{-1} ,
    \label{eq:CRB multi}
\end{equation}  
such that the bound can be rewritten using any positive semidefinite cost matrix $\matriz{R}$ as
\begin{equation}
    \Tr(\matriz{R} \matriz{V})\geq \Tr (\matriz{R} \matriz{F}^{-1})\geq \Tr (\matriz{R} \matriz{Q}^{-1}).
    \label{eq:CRB multi weighted}
\end{equation}

The first inequality in Eq.~\eqref{eq:CRB multi} is again saturable in the asymptotic limit of many trials, so we henceforth deal with the per-trial FI and QFI as defined above. However, the second inequality in Eq.~\eqref{eq:CRB multi} is not always saturable, in that there does not always exist a POVM that can be used to measure the probe state with sufficient sensitivity. Nonetheless, for any practical scenario, this bound is relevant within a factor of 2~\cite{Tsang:2020aa}. Namely, for any linear combination of the covariances that one wishes to minimize, as described by $\matriz{R}$, there exists a POVM satisfying $\Tr(\matriz{R}\matriz{F}^{-1})=\mathsf{f} \Tr(\matriz{R}\matriz{Q}^{-1})$ for a constant $1\leq \mathsf{f}\leq 2$, such that in the asymptotic limit one can attain the minimum weighted uncertainty $\Tr (\matriz{R} \matriz{V})=\mathsf{f}\Tr( \matriz{R} \matriz{Q}^{-1})$. This ensures that maximizing the QFI matrix is still a reasonable goal. The nuances of when the second inequality can and cannot be saturated in Eq.~\eqref{eq:CRB multi} are unique to the domain of multiparameter quantum estimation, under the moniker of parameter compatibility~\cite{Matsumoto:2002ab}. Multiparameter estimation must also properly handle situations in which the QFI matrix is singular~\cite{Goldberg:2021ab}.

A multimode coherent state provides the classical benchmark, which evolves to $\bigotimes_{i=1}^d|\alpha_i \eu^{\iu \phi_i}\rangle$. Since this state is fully separable among the $d$ modes, the covariances of the number operators for different modes all vanish in Eq.~\eqref{eq:QFI multiparam pure}, such that the QFI matrix takes the simple form of a diagonal matrix with components $Q_{ij}=4|\alpha_i|^2\delta_{ij}$. The lower limit on the total variance of all of the phases, equivalent to setting the weight matrix $\matriz{R}$ to identity in Eq.~\eqref{eq:CRB multi weighted}, is
\begin{equation}
    \Tr (\matriz{R}\matriz{V})=\sum_{i=1}^d \mathrm{Var}(\hat{\phi}_i) \geq \frac{1}{4}\sum_{i=1}^d\frac{1}{|\alpha_i|^2}\geq\frac{d^2}{4\bar{n}}
\end{equation} 
for total energy $\bar{n}=\langle \sum_{i=1}^d \hat{n}_i\rangle$.
The invertibility of the diagonal QFI matrix actually implies $\mathrm{Var}(\hat{\phi}_i)\geq 1/4|\alpha_i|^2$ for each and every $i$. 

To achieve the smallest lower bound on the uncertainty sum, each mode must have an equal energy $|\alpha_i|^2=\bar{n}/d$, a property that we will see to recur throughout. Should one be interested in a different cost matrix $\matriz{R}$ that treats one phase uncertainty as more costly than another, the relative weights of the optimal energy distribution for the probe state will change accordingly.

These QFI calculations again presuppose an infinitely strong phase reference for saturating the uncertainty bounds~\cite{Ataman2020,Goldbergetal2020multiphase}. Using a different phase reference $|\beta_i\rangle$ for each state, the variances are each bounded by $4\mathrm{Var}(\hat{\phi}_i)\geq |\alpha_i|^{-2}+|\beta_i|^{-2}$; for a given total energy in the probes $\bar{n}$ and the references $\bar{n}_\beta$, the total uncertainty is bound by
\begin{equation}
    \mathrm{Tr}(\matriz{V})\geq\frac{d^2}{4\bar{n}}+\frac{d^2}{4\bar{n}_\beta}.
\end{equation} 
A more clever technique is to use a single phase reference $|\beta\rangle$ with $\bar{n}_\beta=|\beta|^2$ for a simultaneous measurement of all $d$ phases. When that is performed, the QFI matrix is no longer diagonal and neither is its inverse, with a constant shift $4\matriz{Q}^{-1}=\mathrm{diag}(|\alpha_1|^{-2},|\alpha_2|^{-2},\cdots,|\alpha_d|^{-2})+|\beta|^{-2}$, yielding the minimum total uncertainty
\begin{equation}
    \Tr(\matriz{V} )\geq\frac{d^2}{4\bar{n}}+\frac{d}{4\bar{n}_\beta}.
\end{equation} 
This is our first glimpse into the advantages of simultaneous versus sequential estimation: when the phase reference is not an infinite resource, the uncertainty can be diminished by $(d-1)/4\bar{n}_\beta$ by cleverly using the reference for jointly measuring all phases. This is a small effect when the probe state energies are much smaller than those of the reference; when they are comparable, the total uncertainty can be reduced by a factor of about 2 for large $d$ at the cost of the covariances between the phases getting larger. We will subsequently see that the advantages of simultaneous estimation are more pronounced in the realm of quantum-enhanced probes.

The easiest generalization of NOON states for multiphase estimation is to consider $d$ distinct NOON states, each with a pair of modes used to probe one of the $d$ phases. In such a scenario, the QFI matrix is again diagonal, with elements $Q_{ij}=\delta_{ij}N_i^2$ when the $i$th pair of modes houses a NOON state with total photon number $N_i$. The total energy going through all $d$ phases is now $\bar{n}=\sum_{i=1}^d N_i/2$ such that the total phase uncertainty is lower bounded by
\begin{equation}
    \Tr(\matriz{V})\geq\sum_{i=1}^d\frac{1}{N_i^2}\geq\frac{d^3}{4\bar{n}^2}.
\end{equation} 
Again, the second inequality is achieved when all of NOON states have equal fractions of the total probe energy. Since each of the $d$ states has its own phase reference essentially inbuilt, there is no need to modify this result.

Still, it seems wasteful to use $d$ separate NOON states with a total of $2d$ modes to measure $d$ parameters; we saw with coherent states that a single reference mode could be superior to $d$ reference modes. Consider the ansatz state to be a generalized multimode NOON state of the form~\cite{Humphreys:2013aa}
\begin{equation}
    \label{eq:humphreys}
        |\psi_0\rangle =\beta |N, 0, \ldots, 0 \rangle +\alpha |0, \psi_{\mathrm{NO\cdots ON}}\rangle,
\end{equation}
where
\begin{eqnarray}
        |\psi_{\mathrm{NO\cdots ON}}\rangle & = &  
        \frac{1}{\sqrt{d}} (|N, 0, \ldots, 0\rangle+|0,  N, 0, \ldots, 0\rangle \nonumber \\
        &+ & \cdots+ |0, 0, \ldots, N\rangle).
\end{eqnarray} 
The total energy going through the phases is $\bar{n}=|\alpha|^2N$, with $|\alpha|^2N/d$ going through each phase, while the total energy in the reference mode is $|\beta|^2 N=N(1-|\alpha|^2)$. Such a state has QFI
\begin{equation}
    Q_{ij}=\frac{4N^2|\alpha|^2}{d}\left(\delta_{ij}-\frac{|\alpha|^2}{d}\right);\quad\matriz{Q}^{-1}=\frac{1}{4N^2|\beta|^2}+\frac{d}{4N^2|\alpha|^2} \openone .
\end{equation} The trace yields some factors times $\frac{d}{|\alpha|^2}+\frac{1}{|1-\alpha|^2}$, which is minimized by $|\alpha|^2=\sqrt{d}/(1+\sqrt{d})$, with
\begin{equation}
    \Tr(\matriz{V})\geq \frac{d}{4N^2}\left(\sqrt{d}+1\right)^2=\frac{d^2}{4\bar{n}^2}.
\end{equation} 
This is true when minimizing with respect to $|\alpha|^2$ or strictly minimizing with respect to the probing energy $\bar{n}=|\alpha|^2 N$. The simultaneous estimation scheme thus outperforms the \textit{quantum-enhanced} sequential estimation scheme by a factor of $d$ smaller total uncertainty for the same probe energy. Such a probe is actually optimal when compared to all probes made from superpositions of basis states that each have all $N$ photons contained in only one mode~\cite{Goldberg:2020ad}.

After establishing the optimal probe states, it is necessary to find the optimal measurements to perform on the states after the phases have been imparted. For the coherent states, homodyne measurements saturate the bound, as mentioned above. For the generalized NOON states, an optimal POVM was given in Ref.~\cite{Humphreys:2013aa}, generalized in Ref.~\cite{Pezze:2017aa}, and given with alternative options in Ref.~\cite{Goldberg:2020ab}. One constructs a POVM with $d^2$ elements $\Pi_{n}=|\Upsilon^{(n)}\rangle\langle \Upsilon^{(n)}|$, where each state $|\Upsilon^{(n)}\rangle=\sum_{m}=0 ^d\Upsilon_m^{(n)}|N_m\rangle$ is a superposition of states $|N_m\rangle$ with all $N$ photons in the $m$th mode. The coefficients of the states comprising the POVM elements are given by $\Upsilon_m^{(0)}=1/\sqrt{d+1}$ and, for $n>0$,~\cite{Humphreys:2013aa}
\begin{equation}
    \Upsilon_m^{(n)}=\begin{cases}
        \sqrt{\frac{(n-1)!}{(n+1)!}} &m\leq n-1\\
        -\frac{1}{\sqrt{n+1}}&m=n\\
        0&m>n .
    \end{cases}
\end{equation}

An important consideration in any quantum information protocol is noise. For interferometry, the dominant source of noise is photon loss, which can ruin quantum advantages. Multiphase estimation has the extra possibility of different levels of noise in different arms and there is a simple solution: characterize the loss, then adjust the coefficients in the generalized NOON states of Eq.~\eqref{eq:humphreys} in accordance with the amount of loss per arm. Then, especially in the case of minor photon loss, Heisenberg scaling can optimally be instated~\cite{Yueetal2014,Namkungetal2024njp}.

Other directions in multiphase estimation have been studied. If the multiple phases together represent pieces of one puzzle, such as a network of sensors that are together estimating a global parameter, optimized probe states can be found~\cite{Proctor:2018aa,Ge:2018aa,Eldredgeetal2018}. If one restricts to Gaussian input states that include squeezed coherent states, the simultaneous estimation strategy provides no more than a factor of 2 improvement in total precision~\cite{Gagatsos:2016aa}. If measurements are restricted to photon counting instead of projections onto complicated entangled states, other probe states become optimal~\cite{Rehmanetal2022}. If squeezing and photon counting are allowed for state generation, more exotic probe states can be formed~\cite{Zhangetal2023}. Finally, if one considers global instead of local estimation, where the global scenario has no prior knowledge about the phases being probed, the advantages diminish significantly~\cite{Chesietal2023}.

\section{Platforms and implementations}
 
The developments of the theory in multiphase estimation is still unmatched by experiments. Part of the reasons can be ascribed to the difficulty in achieving the necessary complexity in bulk apparata, due to the number of optical elements to be employed, and the possible intricacy in their arrangement; compare the four schemes for generating generalized NOON states in Ref.~\cite{ZhangChan2018}, where the most feasible one still requires Fock state filtration. An outstanding example is found in Ref.~\cite{Hong:2021aa}, where the authors used a source of polarization-entangled photon pairs to generate generalised NOON states over four spatial modes as a probe for three phases. Remarkably, with their polarization encoding, the authors could perform perform phase-estimation relying on polarization analysis by photon counting, greatly simplifying the setup and surmounting problems of interferometric stability. The authors achieved a total variance $\Tr(\matriz{V})=1.85\pm0.01$, to be compared with 1.5, the value at the QCRB. The authors also emphasize that, although the optimal state $\vert \psi_0\rangle$ in Eq.\eqref{eq:humphreys} could yield an improvement down to 1.4, the optimal measurement would require a more complex arrangement.

The complications of such setups can be remedied by moving to an integrated platform that eases the possibility of cramming components into a small space. Even though this opens new challenges concerning control and reconfigurability, it  constitutes a promising medium for real-life quantum sensors.  To prepare probe states, sources of quantum light based on parametric down-conversion can be used to produce photon pairs, which are then coupled to the on-chip sensor. The simplicity of operating this technical solution has allowed it to be used successfully for multiphase estimation. 

\begin{figure}
    \centering
    \includegraphics[width=0.85\columnwidth]{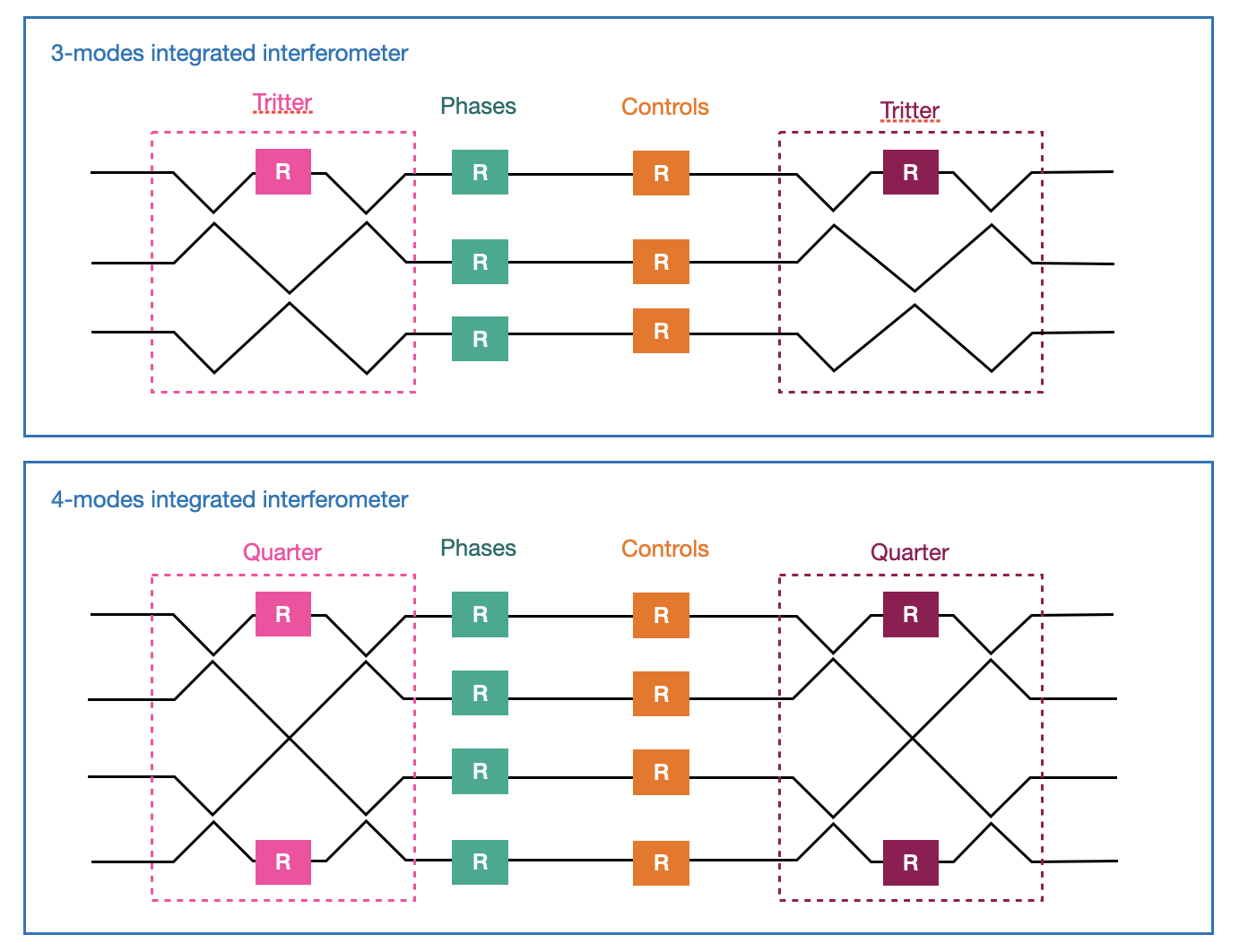}
    \caption{Schemes for three- and four-mode interferometers realized in integrated optical circuits.}
    \label{fig:chip}
\end{figure}

How does multiphase estimation proceed on an integrated photonics device? Waveguides are fabricated on a chip, either by lithography or, more commonly, by laser writing. When two waveguides are brought close, they become evanescently coupled, hence realising a beam splitter with its transmittivity set by the length over which the coupling occurs. Interferometers are thus realised by connecting more such elements, and relative phases are set by fine-tuning the lengths of the waveguides in between beam splitters. Phases can further be controlled by means of thermo-optical shifters relying on the dependence of the refractive index on temperature for small changes; this is controlled by applying a voltage to a resistive device placed above the waveguides. Bringing together several elements, more complex optical devices, such as those shown in Fig.~\ref{fig:chip}, are realised. A tritter is the extension of a beam splitter to three beams and can be realised by composing three beam splitters and a thermo-optic phase shifter; if the evolution is lossless, it results in a unitary transformation of the three-photon input state given by the matrix: ${U}= {U}_{12}(T_3) \mathcal{P}({\theta}) {U}_{23}(T_2)  {U}_{12}(T_1)$, where ${U}_{ij}(T)$ is the unitary transformation associated to the beamsplitter operation between arms $i$ and $j$ with transmittivity $T$ and $\mathcal{P}({\theta})$ is the unitary transformation of the phase shifter which imparts an auxiliary phase $\theta$ on arm 1. This device can serve for both the state preparation and for the measurement hence forming a three-arm interferometer; unknown phases are set on each of the arms and constitute the target parameters, possibly yielding information on temperature, pressure etc. by means of a transduction mechanism. Three more controlled phases are used in order to optimise the measurement conditions and even make them adaptive.

The first tritter is employed to prepare the state, which then propagates along three internal waveguides where thermo-optic phase shifters ensure the possiblity of setting arbitrary relative phases between the three arms, and the second tritter is used at the measurement stage. The chip can be injected with either one, two or three photon states, and it needs to be characterised to carefully account for parameters such as the transmittivities of the beam splitters, the calibration of the phase shifters and other experimental imperfections. The scheme can be extended to four modes, as shown in Fig.~\ref{fig:chip}, and we notice how this now requires setting two phases within the interferometer. These provide more flexibility to the scheme, and, unlike for tritters, the estimation can be optimal~\cite{Pezze:2017aa}. 

The first work on an integrated three-arm interferometer was presented in Ref.~\cite{Polino:2019aa}. The authors implemented the cascaded tritter scheme in Fig.~\ref{fig:chip} in a femotosecond laser written chip. The waveguides were fabricated in order to propagate single photons and photon pairs in the near infrared, 785nm. Photon pairs at these wavelengths are readily produced by means of parametric down-conversion in nonlinear optical crystals. The authors emphasized the importance of characterizing their optical circuit to enable for sensing applications. In particular, they showed the relevance of crosstalk effects in the setting of their thermo-optic shifters, as well as the presence of nonlinear behaviours with the set voltage, due to the temperature dependence of the resistances. The characterisation was carried out by measuring the output probabilities when injecting one photon at time. These yield sufficient information to predict the behaviour when two-photon states are used, also taking into account the level of their indistinguishability $V=0.95\pm0.01$ that sets the maximal contrast of the two-photon interference fringes. These predictions were used on the one hand to investigate the associated FI matrix, and on the other hand to provide an estimator for the two phases: estimation at the CRB limit could be demonstrated when using around 100  repetitions. 

The reason for such a performance needs to be traced to the adoption of a maximum likelihood estimator, which is known to be optimal in the asymptotic limit; i.e., for a large collected sample. However, it offers no guarantee of a similar behaviour for small samples. In this different regime, Bayesian learning protocols, encompassing adaptive steps, have been proposed and demonstrated to provide an effective solution. In Ref.~\cite{Valeri:2020aa} the authors  adopted the same platform as in~\cite{Polino:2019aa}, but only used single photons as probes--these were produced by heralding from the downconversion source. To perform the Bayesian adaptive protocols, two aspects must be taken in due consideration. First, access is needed to a set of control parameters to be varied throughout the estimation in order to maximize the  available posterior information at each step: here, this corresponds to setting the additional thermo-optic phase shifters in the chip. Second, proper numerical methods must be implemented in order to prevent the optimisation from becoming rapidly unmanageable~\cite{Lumino:2018aa,Rambhatla:2020aa}. For this purpose, the Sequential Monte Carlo (SMC) approach was adopted~\cite{Granade:2012aa}: the prior phase distribution $p(\theta)$ is used to extract random values, named {\it particles}, and associate weights to them. These weights are then updated by the Bayesian procedure, providing a way to assess the posterior distribution. Should this become too localised, resampling can be applied. In order to maximise the available information, a target utility function, defined as $\mathcal{U}(\phi)=\Tr [\text{Cov}(\phi)]$ is evaluated on the posterior distribution, setting $\mathsf{R}$ to identity in Eq.~\eqref{eq:CRB multi weighted}. The control phases can then be chosen at each step so that $\mathcal{U}(\phi)$ is minimal. Convergence to the CRB occurs with a scaling with the number $\nu$ of repetitions as $\sim e^{\nu/\tau_N}$, with $\tau_N=5.6$ obtained from the experimental data. Therefore, CRB-limited estimation is reached with approximately 20-30 repetitions. Quantum estimation of multiple phases employing the same adaptive estimation was demonstrated in Ref.~\cite{Valeri:2023aa}, adopting a four arm interferometer, now allowing for the estimation of three independent phases. A two-photon state, exhibiting genuine quantum features in its FI, was used as the probe. The SMC adaptive protocol allowed CRB-limited estimation to be reached with only $\nu\simeq 50$ resources.

These demonstrations have benefitted from an accurate characterisation of the device. In principle, this task can be performed efficiently~\cite{Laing:2012aa,Rahimi:2013aa}; on the other hand, it does not represent a viable solution when deployment is sought outside the research environment, due to the large number of parameters involved. This has prompted the search for alternative approaches, often based on artificial intelligence~\cite{Nolan:2021aa,Fiderer:2021aa}. A first example for multiphase estimation was reported in Ref.\cite{Cimini:2021aa}. The authors sent a photon through an integrated  three-mode interferometer, and collected the experimental frequencies as a function of the applied voltages, $V_1$ and $V_2$, to the thermo-optical elements; the idea is to use these as training data in a feedforward network so that this can learn the association between frequencies and voltages, {\it i.e.} the set phases. In order to resolve ambiguities, the training set also included data taken at $V_1+\Delta V_1$ and  $V_2+\Delta V_2$, where $\Delta V_1, \Delta V_2$ are fixed values chosen so as to optimise the training. A 50$\times$50 grid of voltage values has been deemed sufficient for a satisfactory training. The reliability of the estimation was assessed on 100 random points, generated by a Monte Carlo routine, with more than $80\%$ of them being identified within $0.1$V from the target voltages. 

In Ref.~\cite{Cimini:2023ab}, the authors demonstrate that this same approach can also be complemented with the Bayesian approach: they use the experimental outcomes of their four-mode interferometer in order to train a neural network. Its outcomes, in the form of probabilities for phases conditioned on observed outcomes, are then included in an adaptive Bayes algorithm. Notably, the additional control phases are determined by means of reinforcement learning. This allowed the authors to obtain a complete model-free adaptive estimation protocol, achieving near quantum-limited operation with $\nu \simeq 50-100$ repetitions with a two-photon state, with only a modest excess variance with respect to the ideal case. Numerical tests have identified its origin in using the experimental frequencies to approximate the actual likelihood function of the device. 

A different approach to optimising the settings of the device is based on a variational quantum algorithm that obtains the FI matrix directly from the data~\cite{Cimini:2024aa}. The main difficulty lies in the presence of derivatives in its expression in Eq.~\eqref{eq:fim}. While these could be approximated by a finite-difference ratio, the presence of noise prevents this from achieving satisfactory levels. Therefore, a technique suggested in Ref.~\cite{Cerezo:2021aa} was applied, yielding derivatives by means of numerically stabler differences between experimental frequencies. The availability of an expression for the FI matrix made it possible for the authors of Ref.~\cite{Cimini:2024aa} to set the control phases of their four-mode integrated interferometer to their optimal values.

\section{Conclusion}

Quantum metrology holds great promise for practical, near-term quantum technologies. Experimental progress in sensing is beginning to validate early theoretical predictions and advances in estimation theory. However, a significant theoretical challenge remains: the precise estimation of multiple incompatible observables. In particular, the optimal strategies to define the fundamental precision limits and assess their attainability are not yet fully understood. 

Identifying appropriate platforms for developing new methodologies and benchmarking their performance is a crucial step forward. In this paper, we have reviewed the case of multiphase estimation, which can be implemented in a variety of platforms. New findings in this field are opening exciting avenues for future exploration. Expanding the system's dimensionality could enable investigations into a richer parameter space. The ability to fabricate devices with additional controlled phases provides opportunities to develop and test innovative adaptive protocols or optimize measurement strategies by tailoring the detection operator. Combining these features within the same platform could lead to a new class of optimal protocols, while minimizing resource usage.

\section*{Acknowledgements}

The authors thank V. Cimini and G. Bizzarri for endless but enjoyable discussions on the topic.  This work was supported by the European Commission FET-OPEN-RIA project STORMYTUNE (Grant Agreement No. 899587), Ministero dell'Universit\`a e della Ricerca Dipartimento di Eccellenza 2023-2027, Research Projects of National Relevance (Grant PRIN22-RISQUE-2022T25TR3) and Agencia Estatal de Investigaci\'on (Grant PID2021-127781NB-I00). L.L.S.S. was supported in part by the grant NSF PHY-1748958 to the Kavli Institute for Theoretical Physics (KITP). A.Z.G. acknowledges that
the NRC headquarters is located on the traditional unceded
territory of the Algonquin Anishinaabe and Mohawk people,
as well as support from the NRC Quantum Sensors Challenge
program.

\end{document}